\documentclass[a4paper]{article}

\usepackage[pages=all, color=black, position={current page.south}, placement=bottom, scale=1, opacity=1, vshift=5mm]{background}
\SetBgContents{
	
}      

\usepackage[margin=1in]{geometry} 

\usepackage{amsmath}
\usepackage{amsthm}
\usepackage{amssymb}
\usepackage{bbold}
\usepackage{hyperref}
\hypersetup{
	unicode,
	pdfauthor={Abolfazl Farmanina, Vahid Karimipour},
	pdftitle={Quantum Speed Limits for Implementation of Unitary Transformations},
	pdfsubject={Quantum Speed Limits},
	pdfkeywords={article, template, simple},
	pdfproducer={LaTeX},
	pdfcreator={pdflatex}
}

\usepackage[sort&compress,numbers,square]{natbib}
\theoremstyle{plain}

\theoremstyle{definition}

\usepackage{graphicx}
\graphicspath{{fig/}}
\usepackage{subcaption}
\usepackage{algorithm, algpseudocode} 
\usepackage{mathrsfs} 
\usepackage{lipsum}
\usepackage{physics}
\def\be{\begin{equation}}
	\def\ee{\end{equation}}
\def\ba{\begin{eqnarray}}
	\def\ea{\end{eqnarray}}
\def\lo{\longrightarrow}
\def\h{\hskip 1cm }

\def\la{\langle}
\def\ra{\rangle}
\def\a{\alpha}

\def\ni{\noindent}

\def\bex{\begin{dinglist}{110}\dsquare}
	\def\eee{\end{dinglist}}
\def\bet{\begin{dinglist}{110}\bsquare}
	\def\bfr{\begin{mdframed}[backgroundcolor=blue!20]\vspace{0.5cm}}
		\def\efr{\vspace{0.5cm}\end{mdframed}}
	
	\title{ Quantum Speed Limits for Implementation of Unitary Transformations}
	\author{Abolfazl Farmanian$^1$ \and Vahid Karimipour$^1$}
	
	\date{
		$^1$\small{Deptartment of Physics, Sharif University of Technology, Tehran, Iran} \\%
	}
	
\begin{document}
		\maketitle


\begin{abstract}
	Quantum speed limits are the bounds that define how quickly one quantum state can transform into another. Instead of focusing on the transformation between pairs of states, we 
	provide bounds on the speed limit of quantum evolution by unitary operators in arbitrary dimensions. These do not depend on the initial and final state but depend only on the trace of the  unitary operator that is to be implemented and the gross characteristics (average and variance) of the  energy spectrum of the Hamiltonian which generates this unitary evolution. The bounds that we find can be thought of as the generalization of the Mandelstam-Tamm (MT) and the Margolus-Levitin (ML) bound for state transformations to implementations of unitary operators. We will discuss the application of these bounds in several classes of transformations that are of interest in quantum information processing.
\end{abstract}

\section{Introduction}

\ni Can the universe impose a speed limit, not on the movement of objects, but on the speed of information processing itself? In the recent literature on quantum mechanics and quantum information, this problem is known as the quantum speed limit or QSL for short.  It can be viewed from several different perspecstives. The first one which is mainly conceptual, understands it as a more precise formulation of the Heisenberg Energy-Time uncertainty relation\cite{aharonov_1961_time} which has been known since the early days of quantum mechanics. However, with the upsurge of quantum technology, this problem has gained renewed attention as a practical matter of great interest. Given an amount of energy and a pair of states, what is the shortest time for transforming one element of the pair to the other? This point of view is important for speed of quantum computation  \cite{lloyd_2000_ultimate}.  The same question  can be asked in a different form: Given an amount of time, what is the least amount of energy that we need for the evolution of a given state to another one\cite{bekenstein_1981_energy}. This point of view emphasizes the equally important problem of energy consumption in quantum computation. Regardless of these points of view, the first attempt for formulating the quantum speed limit goes back to Mandlestam and Tamm \cite{mandelstam_1991_the} and then to Margolus and Levitin \cite{margolus_1998_the}, the combined result of whom can be formulated as follows: Given an initial state $|\psi_0\ra$, and final state $|\psi_\tau\ra$, such that $\la \psi_0|\psi_\tau\ra=0$, the time needed for the above evolution is subject to the following bound \cite{levitin_2009_fundamental}: 
\be\label{tau1}
T \geq  T_{QSL}=Max \Big\{\frac{\hbar \pi}{2\Delta E},\frac{\hbar \pi}{2E}\Big\},
\ee
where $H$ is the Hamiltonian  which performs this evolution, $ E=\la \psi_0|H|\psi_0\ra - E_0$, is the average energy about the ground state energy $E_0$, and $\Delta E=\la \psi_0|H^2|\psi_0\ra-\la \psi_0|H|\psi_0\ra^2$ is the variance of energy in the initial state. This relation was then generalized to the case where the initial and final states are no longer orthogonal \cite{pfeifer_1993_how,giovannetti_2003_quantum,niklashrnedal_2023_margoluslevitin}. 
	 Since then, various aspects of quantum speed limit have been extensively studied. These works include QLS bound for mixed states and nonunitary evolutions\cite{taddei_2013_quantum,delcampo_2013_quantum,deffner_2013_energytime,campaioli_2018_tightening,funo_2019_speed,brody_2019_evolution}, 		
	 relation between  entanglement and QLS \cite{giovannetti_2003_the,josepbatle_2005_connection,rudnicki_2021_quantum}, speed limits for the evolution of observables in the Heisenberg picture \cite{mohan_2022_quantum},  the study of speed limit for a bounded energy spectrum\cite{ness_2022_quantum}, and other interesting aspects of quantum speed limit
	\cite{zhang_2023_quantum,ashrafnaderzadehostad_2024_optimal,huang_2023_path,niklashrnedal_2022_ultimate,vu_2023_topological,poggi_2021_diverging,fadel_2021_timeenergy,gessner_2018_statistical,diegopaivapires_2016_generalized}. For a review on this subject see \cite{frey_2016_quantum,deffner_2017_quantum}.\\ 
	
\ni	In this regard we particularly mention the works \cite{poggi_2019_geometric} and  \cite{aifer_2022_from}, where quantum speed limits are investigated for unitary evolutions and \cite{moeinnaseri_2024_quantum} where the quantum speed limit for a change of basis between Mutually Unbiased Bases (MUB's) is studied. In \cite{poggi_2019_geometric}, starting from the Fubini-Study distance between states, a time-energy bound is derived, known as the Anandan-Aharonov relation. Using this distance, the author proceeds to define a distance between unitary operators and derive  a time-energy bound for unitary evolution in the form 

		\be\label{poggi1}
	T\geq \frac{\sqrt{N}\arccos(\frac{1}{N}|\tr(U)|)}{\sqrt{\tr(H^2)}}=\frac{\arccos(\frac{1}{N}|\tr(U)|)}{\sqrt{\frac{\sum_{k=1}^N E_k^2}{N}}}.
	\ee
The problem with this interesting bound is that an arbitrary shift in all the energy levels of the Hamiltonian, while leading to the same unitary evolution, yields a completely different time-energy bound as is evident from  (\ref{poggi1}).\\

\ni Starting again from time-energy bounds for evolution of states, the authors of \cite{aifer_2022_from}, try to remove the dependence on the energy of the initial and final states in (\ref{tau1}) and replace it with the energy spectrum of the Hamiltonian. To this end they invoke Popoviciu inequality $\Delta E\leq (E_{max}-E_{min})/2$ \cite{bhatia_2000_a}  (where $E_{max}$ is the highest energy level and $E_{min}$ is the energy of ground state) and arrive at  
\be
\tau\geq \tau_{QSL}=\frac{2\hbar \theta}{ E_{max}-E_{min}}.
\ee
\ni The new bound now depends on the width of the spectrum of the Hamiltonian  evolving the state. However, the bound still depends on $\theta$, the angle between the initial and final states. To remedy this, the authors of \cite{aifer_2022_from} define $\theta$ for qubits to be  the maximum arc angle on bloch sphere that the specified unitary can transform a state. In this way they are able to find time-energy bounds for unitaries acting on $n$ qubits, i.e. for $2^n-$ dimensional unitaries, not for general ones.  Moreover, their definition for calculating this angle $\theta$ for an $n-$ qubit unitary does not lead to a straightforward calculation. \\

\ni Finally in \cite{moeinnaseri_2024_quantum}, the authors take a different route and instead of transformation between two  single states, consider transformation between two special sets of states, namely those sets which are mutually unbiased with respect to each other. In other words, they consider transformations between Mutually Unbiased Bases or MUB's for short. 
That is, they ask what is the quantum speed limit if we want to convert a complete basis $\{|k\ra,\  k=1\cdots d\}$ to another one $\{|e_k\ra, k=1\cdots d\}$ which is a Mutually Unbiased Basis (MUB) with respect to the original one. They proceed to find quantum speed limits for change of basis for  qubit and qutrit bases and conjecture  general bounds for general dimensions. ‌Besides the restriction to certain  classes of unitary transformations, a problem with this approach is that they ignore change of relative  phases in evolution between MUB bases, i.e. they allow that an arbitrary state $|\psi\ra=\a|0\ra+b|1\ra$  be converted to $|\psi'\ra=\a|+\ra + e^{i\phi} b|-\ra$.  This is inherent in their approach which considers unitaries of the form $U=\sum_{k} e^{i\phi_k}|e_k\ra\la k|$, where the phases $e^{i\phi}$ are to meant to be eliminated by another unitary of the form $V=\sum_{k} e^{-i\phi_k}|e_k\ra\la e_k|$.
Therefore, if we have fixed a reference frame for phases, then the energy-time constraint for the diagonal unitary operator $V$ should also be taken into account. Needless to say, one can redefine the basis states to remove the phases, however this will become problematic if we consider consecutive unitary gates acting in a circuit. \\

\ni In this article, we follow a different approach, namely we derive quantum speed limits for unitary evolution from the very beginning, without resorting to speed limits for transformation of pairs of states. In contrast with \cite{poggi_2019_geometric}, our bound is invariant under a shift of energy levels and in contrast with \cite{moeinnaseri_2024_quantum}, we consider transformation of a basis $B=\{|i\ra,\   i=0,1,\cdots, N-1\}$ to another arbitrary one $B'=\{|e_i\ra, i=0,1,\cdots, N-1\}$, be it MUB or not with respect to the original one, without producing any extra relative phase.\\

\ni {\bf Remark:} One may argue that given a unitary operator $U=e^{-iHT}$, one can first calculate the eigenvalues $\lambda_k$ of $U$  and then proceed to extract the energies via the relation $\lambda_k(U)=e^{-iE_kT}$ from which a time-energy bound can be established \cite{juyeongyhm_2024_minimal}. There are however two major problems with this approach. The first obvious one is that finding the logarithm of a unitary, specially in high dimensions is a highly nontrivial problem. The second more important one is that, from the relation $\lambda_k(U)=e^{-iE_kT}$, each energy level or more precisely each $E_kT$ is obtained only modulo an arbitrary multiple of $2\pi$ (i.e. $E_kT$ and $E_kT+2\pi n_k$ where $n_k$ is an integer lead to the same eigenvalue of $\lambda_k(U)$). Of course one can restrict all the $E_kT$'s to lie in the interval $(-\pi, \pi)$ for obtaining the minimum average energy. However, even then,  an ambiguity in the correct time-energy relations remains. This is due to the fact that from the spectrum of $U$ arranged on the unit circle in the complex plane, in the form $\lambda_k(U)=e^{-iE_kT}$,  each of the energies $E_k$ can be shifted to be the ground state. This obviously affects the average energy above the ground state according to the definition (\ref{n0}).\\

\ni The time-energy bound that we propose circumvents these two problems. It only needs a knowledge of $tr(U)$, is free from the phase ambiguity of taking the logarithm of $U$ and depends solely on  $tr(U)$. This bound in fact generalizes in a simple way the Mandelstamm-Taam and Margolous-Levitin bounds to implementation of unitaries and can be useful in  design of quantum circuits and other quantum information processing tasks. \\

\ni The structure of this paper is as follows:  in section \ref{main}, we state our main result in the form a proposition.  This proposition is then proved in section \ref{proof}. In section \ref{ex}  we calculate the bounds for several classes of unitaries which are important for quantum computation algorithms. Finally in section \ref{exp}, we discuss the tightness of these bounds for qubit and qutrit cases. we conclude the paper with a discussion. \\

\ni {\Large{{\bf Notations:}}}
Throughout the paper we take $\hbar=1$ and consider a system which is represented in a finite dimensional Hilbert space of dimension $N$ with discrete energy eigenvalues are ordered as $\{E_0\leq \cdots \leq E_i\leq \cdots E_{N-1}\}$. Thus some of the energy eigenvalues may be degenerate. Therefore $E_0=E_{min}$ is the minimum energy eigenvalue and $E_{N-1}=E_{max}$ is the maximum one . The average energy above the ground state is denoted by
\be\label{n0}
E:=\frac{\sum_{k}E_k}{N}-E_0
\ee
and its variance by 
\be 
(\Delta E)^2=\frac{1}{N}\sum_{k=0}^{N-1} (E_k-\overline{E})^2.
\ee
The width of the spectrum is also denoted by $\delta E$:
\be
\delta E:=E_{max}-E_{min}.
\ee 
\section{Main results}\label{main}
Our main result is the following proposition which we state below and prove in the next section.\\

\ni {\bf Proposition:} To affect an $N-$ dimensional unitary gate $U=e^{-iHT}$, the following quantum speed limit holds:\\

\be
\label{b1}
T\geq  Max\left(\frac{\pi}{2E} \Big(1-\frac{|tr(U)|}{N}\sqrt{1+\frac{4}{\pi^2}}\Big)\ ,\ \frac{1}{\Delta E}\sqrt{1- \frac{|tr(U)|^2}{N^2} }\right)
\ee2

\ni This bound is the analogue of combined Mandlestam-Tamm(MT) and the Margolus-Levitin (ML) bound \cite{mandelstam_1991_the,margolus_1998_the}.\\  

\ni Note that this bound depends only on $|tr(U)|$. This reflects the equivalence of the gates $U$ and $e^{i\phi}U$, the latter being equivalent to a shift of all energies. The bound is also the same for the gates $U$ and $VUV^\dagger$, where $V$ is a unitary which changes the bases of the Hilbert space, in which $U$ is expressed as a matrix. \\

\ni {\bf Corollary:} From the main proposition also follows a bound which depends on the width of the spectrum and is written as

\be\label{corr} T\geq Max \left( \frac{\pi}{\delta E}\Big(1-\frac{|tr(U)|}{N}\sqrt{1+\frac{4}{\pi^2}}\Big) \ , \ \frac{2}{\delta E} {\sqrt{1- \frac{|tr(U)|^2}{N^2}  }}\ \ \right),
\ee
where $\delta E=E_{max}-E_{min}$ is the width of the spectrum.

\section{ Proof of the proposition }\label{proof}
\ni Let $U$ be the Unitary and $H$ be the Hamiltonian with energy levels in the set $$\{ E_0\leq \cdots E_i\cdots \leq E_{N-1}, \ \ i=0,1,\cdots N-1 \},$$ which is responsible for transformation of the basis $B_0=\{|0\ra,|1\ra, \cdots,|N-1\ra\}$ to a  basis $B_1 =\{|e_i\ra, i=0,1,\cdots N-1\}$. The unitary operator corresponding to this evolution is given by $U=e^{-iHT}$.  \\

\ni Consider the well-known identity
\begin{equation} \label{Inq1}
	x \geq \frac{\pi}{2}(1-\cos x) - \sin x, \ \ \  \ \ \ \forall \ x\geq 0.
\end{equation}
Let $x=(E_{k}-E_0)T$ in (\ref{Inq1}) and sum over $k$ to find :
\be
\label{sumE_1}
\sum_i (E_k -E_0) T \geq  \frac{N\pi}{2} - \frac{\pi}{2} \sum_k \cos (E_k-E_0)T - \sum_k \sin (E_k-E_0)T\ee
With a little rearrangement, equation (\ref{sumE_1}) can be rewritten as

 \be		
\sum_i (E_k -E_0) T 	\geq  N \frac{\pi}{2}-(\sum_k \cos E_kT)(\frac{\pi}{2}\cos E_0T-\sin E_0T)-(\sum_k \sin E_kT)(\frac{\pi}{2}\sin E_0T+\cos E_0T).
\ee
Noting that 
\be
tr(U)=\sum_k e^{-iE_kT}=\sum_k \cos E_kT - i\sum_k \sin E_kT := |tr(U)|(\cos \theta-i\sin\theta)
\ee
for some angle $\theta$, and equivalently that
\be\label{ucos} \sum_k \cos E_kT:=|tr(U)|\cos\theta,\h \sum_k\sin E_kT:=|tr(U)|\sin\theta.\ee
In view of (\ref{ucos})  we find 
\be\label{emin}
N(\overline{E}-E_{min}) T \geq   N \frac{\pi}{2}-|tr(U)|\big(\frac{\pi}{2}\cos(\theta-E_0T)-\sin(\theta-E_0T)\big).
\ee
Since the last trigonometic expression is always less than $\sqrt{1+\frac{\pi^2}{4}}$, this leads to 
\be\label{emin1}
N(\overline{E}-E_{min}) T \geq   N \frac{\pi}{2}-|tr(U)|\big(\sqrt{1+\frac{\pi^2}{4}}\big).
\ee
or in view of the definition $E=\overline{E}-E_{min}$, as 
\be
\label{inqE}
E T \geq  \frac{\pi}{2} -\frac{|tr(U)|}{N}\big(\frac{\pi}{2}\cos(\theta-E_0T)-\sin(\theta-E_0 T)\big). 
\ee
and hence we find  
\be\label{ineq1}
T\geq  \frac{\pi}{2E} \Big(1-\frac{|tr(U)|\sqrt{1+\frac{4}{\pi^2}}}{N}\Big).
\ee
This is the analogue of the Margolus-Levitin bound. To prove the analogue of the Mandelstam-Tamm bound, we use the identitiy 
\begin{equation} \label{x2}
	x^2 \geq 2(1-\cos x),\h \forall\ \ x.
\end{equation}
Setting x = $(E_k - E_l)T$ in (\ref{x2}) and summing over all $k$ and $l$ we have:
\begin{equation}
	\sum_{k,l} (E_k -E_l )^2T^2 \geq 2 	\sum_{k,l}  (1-\cos((E_k -E_l )T)  ) \label{delatE1}
\end{equation}
We note that 
\be \sum_{k,l=1}^N (E_k-E_l)^2=2N(\sum_k E_k^2)-2(\sum_k E_k)^2=2N^2\Delta E^2\ee
and
\begin{equation}
	|tr (U) | ^2 = \sum_{k}e^{-iE_kT}\sum_l e^{iE_lT}=\sum_{k,l}e^{-i(E_k-E_l)T}=\sum_{k,l}  \cos(E_k -E_l )T,
\end{equation}
where we have used the odd parity of $\sin(E_k-E_l)T$.
Inserting these two last equalities in (\ref{delatE1}), we find 
\begin{equation}
	\Delta E^2  T^2 \geq  (1- \frac{|tr(U)|^2}{N^2}  ) \label{delatE3}
\end{equation}
or
\begin{equation}	\label{DeltaE}
	T \geq \frac{\sqrt{(1- \frac{|tr(U)|^2}{N^2}  )}}{\Delta E} 
\end{equation}
Combination of the two inequalities  (\ref{ineq1}) and (\ref{DeltaE}), proves the proposition. \\

\ni {\bf Proof of the corollary:}\\

\ni If we use Popoviciu identity \cite{bhatia_2000_a}, namely,  $2\Delta E \leq E_{max} - E_{min}=\delta E$, we can also obtain a new bound from (\ref{DeltaE}) which is based on the width of the spectrum, that is
\begin{equation} \label{Emax1}
	T \geq \frac{2\sqrt{(1- \frac{|tr(U)|^2}{N^2}  )}}{\delta E} .
\end{equation}

\ni Alternatively, we can prove an analogue of the Dual Margolus-Levitin bound \cite{ness_2022_quantum}by taking $x=(E_{max}-E_k)T$ in (\ref{Inq1}) and exactly follow the same steps to arrive at 
\be \label{dualmeanE}
N(E_{max}-\overline{E}) T \geq   N \frac{\pi}{2}-|tr(U)|\big(\sqrt{1+\frac{\pi^2}{4}}\big).
\ee
or 
\be
\label{ineqdual}
T\geq  \frac{\pi}{2(E_{max}-\overline{E})} \Big(1-\frac{|tr(U)|\sqrt{1+\frac{4}{\pi^2}}}{N}\Big).
\ee
Summing (\ref{dualmeanE}) with (\ref{emin1}), we find
\be
N(E_{max}-E_{min}) T \geq   N {\pi}-|tr(U)|\big(\sqrt{4+\pi^2}\big),
\ee
or
\be\label{diff2}
T\geq  \frac{\pi}{\delta E} \Big(1-\frac{|tr(U)|}{N}\sqrt{1+\frac{4}{\pi^2}}\Big).
\ee
Combining the two bounds (\ref{diff2}) and (\ref{Emax1}), proves the corollary (\ref{corr}).

\section{Examples }\label{ex}
We now provide several case scenarios of applying the bounds on the unitary gates.  The examples which are chosen are the ones which have practical applications in quantum information and computation theory.

\subsection{Transformation between Mutually Unbiased Bases or MUB's}\label{MUB}
Let $U$ be a transformation which maps $B_0=\{|0\ra, |1\ra,\cdots |N-1\ra \}$ to a mutually unbiased bases $B_1=\{|e_0\ra, |e_1\ra, \cdots |e_{N-1}\ra \}$, where  $|<k|e_l>|^2 = \frac{1}{N}\ \ \forall\ \ k, \ l$. This implies that:
$
| tr(U) | \leq \sqrt{N} \label{MUBinq} 
$
and hence the bound (\ref{b1}) reads:
\be
T\geq Max\left(\frac{\pi}{2E}( 1 - \frac{1}{\sqrt{N}}\sqrt{1+\frac{4}{\pi^2}})\ ,\  \frac{\sqrt{1-\frac{1}{N}}}{\Delta E}\right) \label{MUBinq1}
\ee
In \cite{moeinnaseri_2024_quantum}, a bound is obtained for transformation between $MUB$'s which can be expressed as follows:
\be\label{moin}  T\geq \begin{cases} \frac{2\pi}{9 E} & \text{for\ \  N=3}\\ \frac{\pi(N-1)}{4N E}& \text{for}  \  N\geq 4\end{cases}\ee
It is seen that for all dimensions $N\geq 4$, our bound is tighter than the bound (\ref{moin})  obtained in \cite{moeinnaseri_2024_quantum}. Specially for very large dimensions the right-hand side of our bound converges to $ \frac{\pi}{2}$ while the one given in (\ref{moin}) converges to $\frac{\pi}{4}$.
Note also that for every $N$-qubit system, there is an explicit example, namely $U=H^{\otimes N}$ \cite{moeinnaseri_2024_quantum} which converts a  basis to a MUB basis in $T=\frac{\pi}{2E}$  This shows that for large $N$, the  bound in (\ref{MUBinq1}) becomes tight. \\

\ni As a special case of transformation between MUB's, the Fourier Transform  is defined as $$F=\frac{1}{\sqrt{N}}\sum_{k,l=0}^{N-1}\omega^{kl}|k\ra\la l|,$$ where $\omega=e^{\frac{2\pi i}{N}}$.  A closed form for the trace of $F$ is given in the form of  general quadratic Gauss sum  \cite{berndt_1981_the}. This is given by
\be
|tr(F)| =|\frac{1}{\sqrt{N}}\sum_{k=0}^{N-1}\omega^{k^2}|=\begin{cases} \sqrt{2}& {\rm if }\ N=0\ \ \ \ {\rm mod} 4\\
	1& {\rm if }\ N=1,3\ {\rm mod} 4\\
	0& {\rm if }\ N=2\ \ \ \  {\rm mod} 4\\
\end{cases}.
\ee
One can insert these values in (\ref{b1}) and (\ref{corr}) and obtain time-energy bounds for Fourier transform in any dimension. \\
\subsection{The permutation operator}
\ni As another special case, we can consider a permutation operator $P$, which permutes the basis states in such a way that a number $m$  of basis states are kept fixed. In this case,   the operator $P$ is real and $Tr(P)=m$ leading to 
\begin{equation}
	T\geq Max\left(\frac{\pi}{2E}(1-\frac{m}{N}\sqrt{1+\frac{4}{\pi^2}})\ ,\ \frac{1}{\Delta E}\sqrt{1+\frac{m^2}{N^2}}\right).
\end{equation}
In passing we note that the bound given in proposition 6 of \cite{moeinnaseri_2024_quantum} applies only to shift operator $U|n\ra=|(n+1)\ mod\ N\ra$ and not to any permutation. This is evident from the proof of this proposition in \cite{moeinnaseri_2024_quantum} which hinges upon a form of eigenvalues specific to the shift operator and not to any permutation operator. 

\subsection{The Grover operator}
The Grover operator is defined as $G=(2|s\ra\la s|-I_N)(I_N-2|\omega\ra\la \omega|)$, where $I_N$ is the $N$-dimensional identity operator, $|\omega\ra$ is the single state which is to be found and $|s\ra=\frac{1}{\sqrt{N}}(\sum_{x=1}^N|x\ra)$. We find
$
|tr(G)|=N-4+\frac{4}{N}.
$
In this case, we find from (\ref{DeltaE}) a time-energy bounds for the Grover operator in any dimension  

\be
T\geq \frac{\sqrt{1-(1-\frac{4}{N}+\frac{4}{N^2})^2}}{\Delta E}\approx \frac{2\sqrt{2}}{\Delta E \sqrt{N}},
\ee 
where the last expression is given for large $N$.

\section{An examination  of the bounds in low dimensions}\label{exp}
To evaluate our bounds and see how tight they are, we examine them in low dimensions, where explicit calculations of the eigenvalues of the unitary operator is possible. We directly determine the Energy-Time relation by explicit calculation and determine how well our bounds are compared with them. Let us consider the qubit and the qutrit cases separately.  
\subsection{The qubit case:}
A general unitary operator for qubit systems is  of the form :
\begin{equation}\label{U2}
	U=e^{i\phi} \begin{pmatrix}e^{i\alpha}\cos\theta&e^{i\beta}\sin\theta \\ -e^{-i\beta}\sin\theta & e^{-i\alpha}\cos\theta \end{pmatrix}.
\end{equation}
We assume that this  unitary operator is affected by a Hamiltonian $H$ which is constant in time and therefore $U=e^{-iHT}$. To find the energy levels of the Hamiltonian, we find the eigenvalues of the operator $U$, which turn out to be: 
\begin{equation}
	\lambda_{\pm} =e^{i\phi}\Big( \cos \theta\cos\alpha \pm i\sqrt{1-\cos^2\theta\cos^2\alpha}\Big )
\end{equation}

\begin{figure}[H]
	\centering
	\includegraphics[width=12cm]{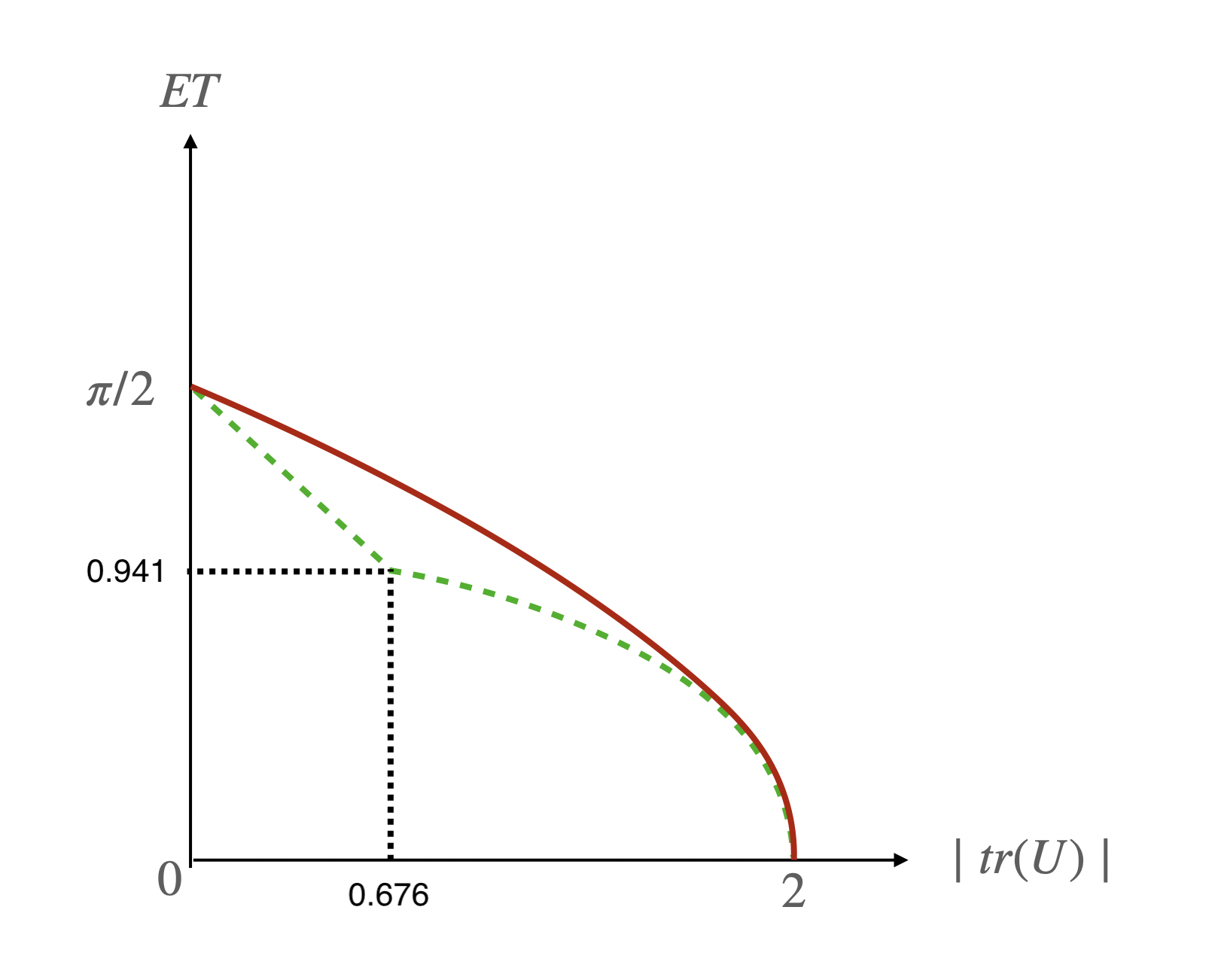}\vspace{0 cm}
	\caption{ (Color online) The exact time versus $|\tr(U)|$ for enacting a unitary operator on a qubit system (solid line), compared with the bound (\ref{b1}) (the dashed line). For qubits, $\Delta E=E=\frac{E_2-E_1}{2}$ and hence both bounds in (\ref{b1}) can be expressed as $ET$ versus $|\tr(U)|$. }
	\label{qubit}
\end{figure}

\ni In view of the relation $U=e^{-iHT}$, this gives the following relation:
\begin{equation}
	(E_2 -E_1)T =2 \cos^{-1}(|\cos\theta\cos\alpha|) 
\end{equation}
where $E_1\leq E_2$ are the energies of the Hamiltonian. Since from (\ref{U2}), $|tr(U)|=2|\cos\theta\cos\a|$ and since for a two-level system, we have $E=\frac{E_1+E_2}{2}-E_1=\frac{E_2-E_1}{2}$, this can be written as 
\begin{equation}\label{qubitinq}
	T = \frac{\cos^{-1}(\frac{|tr(U)|}{2})}{E}.
\end{equation}
This relation gives the exact time for enacting the operator $U=e^{-iHT}$ on a two-level system. Figure (\ref{qubit}) compares our bound (\ref{b1}) with the actual time.

\begin{figure}[H]
	\centering
	\includegraphics[width=12cm]{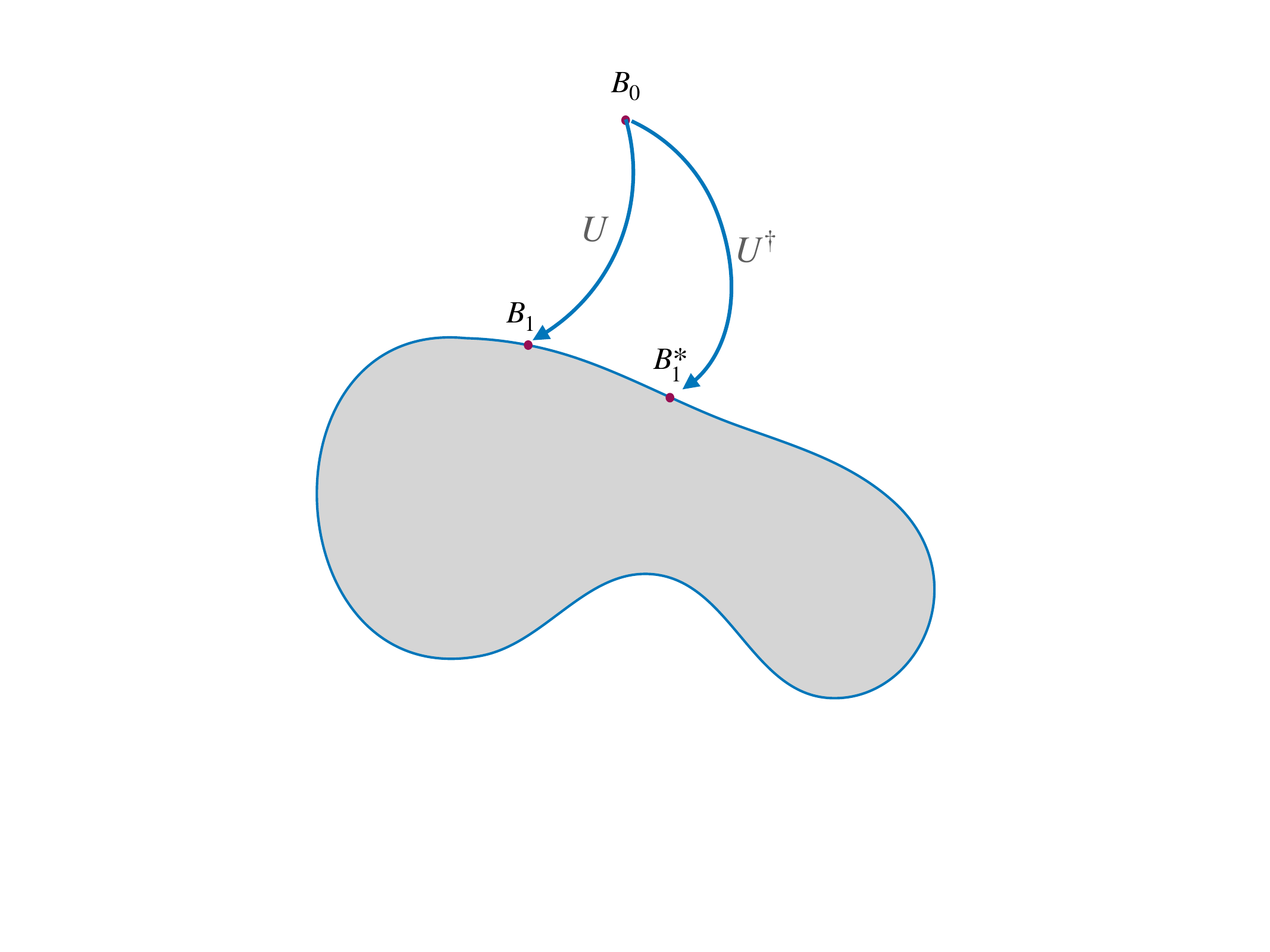}\vspace{-2 cm}
	\caption{The basis $B_0=\{|0\ra,|1\ra\}$ can be transformed to a basis in the set of all MUB's (shaded region). With a given energy, transforming to the  basis $B_1^*=\{ |e_0\ra,|e_1\ra\}$  as given in (\ref{b1*}) needs the shortest amount of time. }
	\label{UU}
\end{figure}

\ni {\bf The energy time relation for transformation between qubit MUB's:}
Equation (\ref{qubitinq}) gives the energy time relation for any qubit unitary operator. Note that this is an equality, obtained by exact solution of the spectrum of the Hamiltonian. For a unitary operator which transforms the basis $B_0=\{|0\ra, |1\ra\}$ to a basis which is MUB with respect to $B_0$, we should set $\theta = \frac{\pi}{4}$.  Therefore equation (\ref{qubitinq}) gives the relation 
$T=\frac{\cos^{-1}(\frac{|\cos(\alpha)|}{\sqrt{2}})}{E}$
which takes its minimum value for $\a=0 $ and $\a=\pi$. This corresponds to the following MUB basis with respect to $B_1$: 
\be\label{b1*}
B^*_1=\Big\{|e_0\ra=\frac{1}{\sqrt{2}}\begin{pmatrix}1\\  e^{i\phi}\end{pmatrix}\  \ , \  \ |e_1\ra=\frac{1}{\sqrt{2}}\begin{pmatrix}- e^{-i\phi}\\ 1\end{pmatrix}\,  \Big\}.
\ee
\ni What is the meaning of the minimization carried over $\a$ or $|tr(U)|$?  It means that among the set of all MUB bases available, the one  which can be reached in the shortest possible time with the  given amount of energy $E_2-E_1$ is the MUB with $\a=0$ or $\alpha =\pi$.   Figure (\ref{UU}) illustrates this point. 

\subsection{The qutrit case}
Unlike the qubit case, the exact determination of an energy-time relation for general three-dimensional unitaries is not feasible. This is due to the large number of parameters that such a unitary has which renders exact determination of its eigenvalues and the corresponding Hamiltonian and hence an exact energy time relation is intractable or at least not illuminating.  We can do this for a subset of such unitaries, namely, the ones which perform transformations between MUB bases. This also enables us to compare our results with  \cite{moeinnaseri_2024_quantum}. Therefore we restrict ourselves to those unitaries which transform the basis state $B_0=\{|0\ra, |1\ra, |2\ra\}$ to a basis $B_1=\{|e_0\ra, |e_1\ra, |e_2\ra \}$ which is MUB with respect  $B_0$. A unitary operator which performs this transformation, without inducing extra phases is given by  $U=\sum_{i=0}^2 |e_i\ra\la i|$. It has been shown that that the most general unitary operator performing this task falls into one of the following two classes \cite{moeinnaseri_2024_quantum}:
\begin{equation}
	U_1 = \frac{1}{\sqrt{3}} \begin{pmatrix}
		e^{i\phi_1} & 	 e^{i(\phi_1 - \alpha)}& 	 e^{i(\phi_1-\beta)}\\
		e^{i\phi_2}& 	\omega^2 e^{i(\phi_2 -\alpha)} & 	\omega  e^{i(\phi_2 - \beta )} \\
		e^{i\phi_3} &	 \omega e^{i(\phi_3 - \alpha )}&	\omega^2  e^{i(\phi_3 - \beta )}
	\end{pmatrix} \label{11}
\end{equation}
and
\begin{equation}
	U_2 = \frac{1}{\sqrt{3}} \begin{pmatrix}
		e^{i\phi_1} & 	 e^{i(\phi_1 - \alpha)}& 	 e^{i(\phi_1-\beta)}\\
		e^{i\phi_2}& 	 \omega e^{i(\phi_2 -\alpha)} & 	 \omega^2 e^{i(\phi_2 - \beta )} \\
		e^{i\phi_3} &	\omega^2  e^{i(\phi_3 - \alpha )}&	 \omega e^{i(\phi_3 - \beta )}
	\end{pmatrix} \label{12}
\end{equation}
where $\omega=e^{\frac{2\pi i}{3}}$.
The set of MUB's is thus a continuous five dimensional manifold with local coordinates $\{\phi_1, \phi_2, \phi_3, \a, \beta\}$ . Let us denote this manifold by ${\cal M}$. 
One can remove the phases $\phi_1, \phi_2$ and $\phi_3$ by first transforming $U\lo e^{-i\phi_1 }U$ and then transforming $U\lo VUV^\dagger $, where $V=diag(1, e^{i(\phi_1-\phi_2)}, e^{i(\phi_1-\phi_3)})$. These two transformations still  retain the property of MUB-ness, i.e. map the manifold of MUB bases into itself. 
The first transformation is affected by a shift of all energy levels of the Hamiltonian $H\lo H+\frac{\phi_1}{T}$ and the second one is affected by a conjugation of the Hamiltonian $H\lo V HV^{\dagger}$. Both of these Hamiltonians have the same energy characteristics $ E$ and $\Delta E$ as the original one. The net result is that if the Hamiltonian $H$ brings the basis $B_0$ to a point $B_1$ in the manifold ${\cal M}$, another Hamiltonian with the same energy characteristics, can also reach another point of ${\cal M}$ at the same time, i.e. under the same time-energy bound. 
\begin{figure}[H]
	\centering
	\includegraphics[width=12cm]{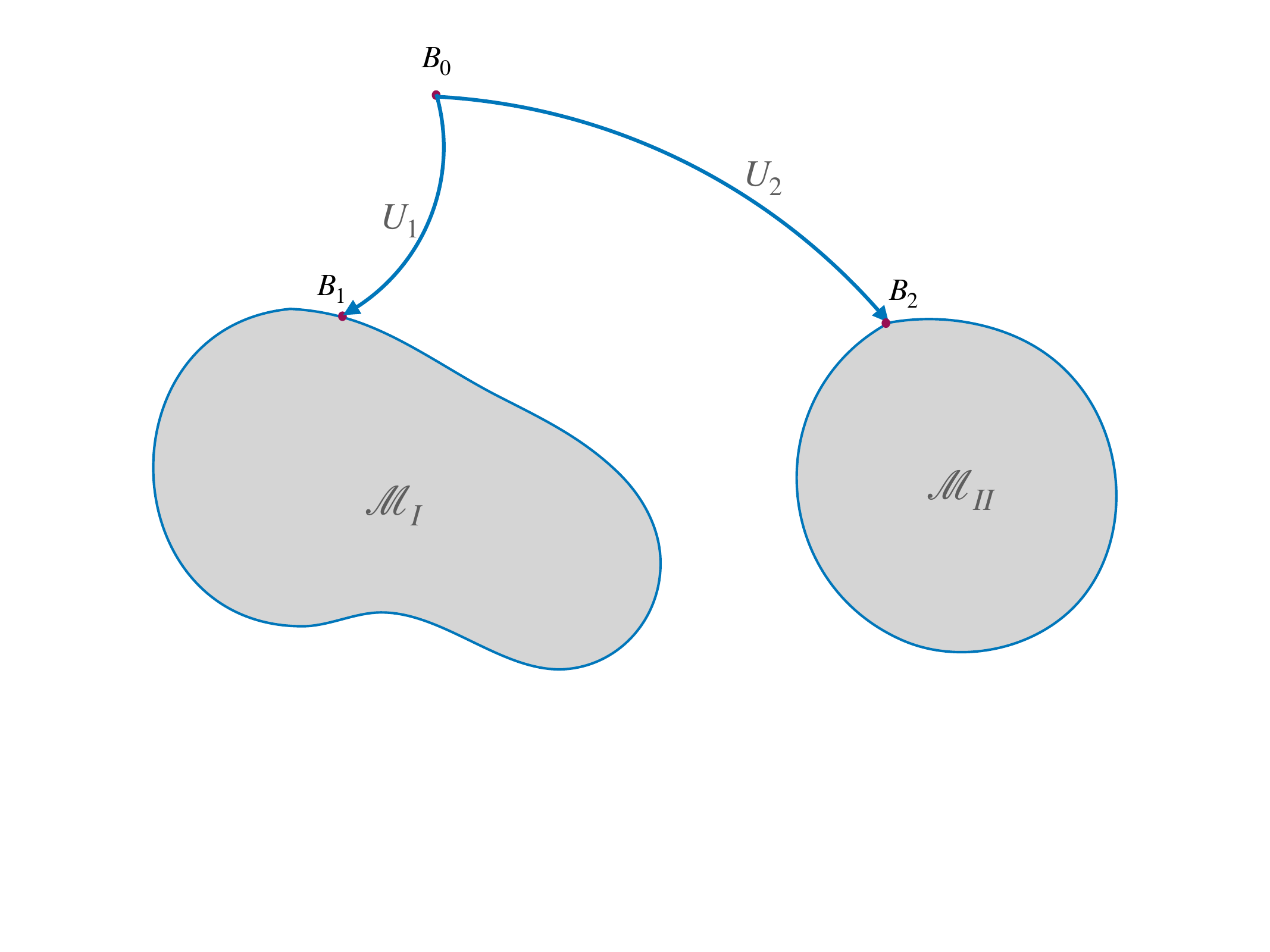}\vspace{-2 cm}
	\caption{The manifold of MUB bases for qutrits has two disjoint parts. With a given amount of energy, the basis $B_0=\{|0\ra,|1\ra, |2\ra\}$ can be transformed in the shortest time  to a MUB basis $B_1 \in {\cal M}_I$ or to a MUB basis  $B_2\in {\cal M}_{II}$.  The energy-time bound for the two sets are different.   }
	\label{UUU}
\end{figure}

\begin{figure}[H]
	\centering
	\includegraphics[width=14cm]{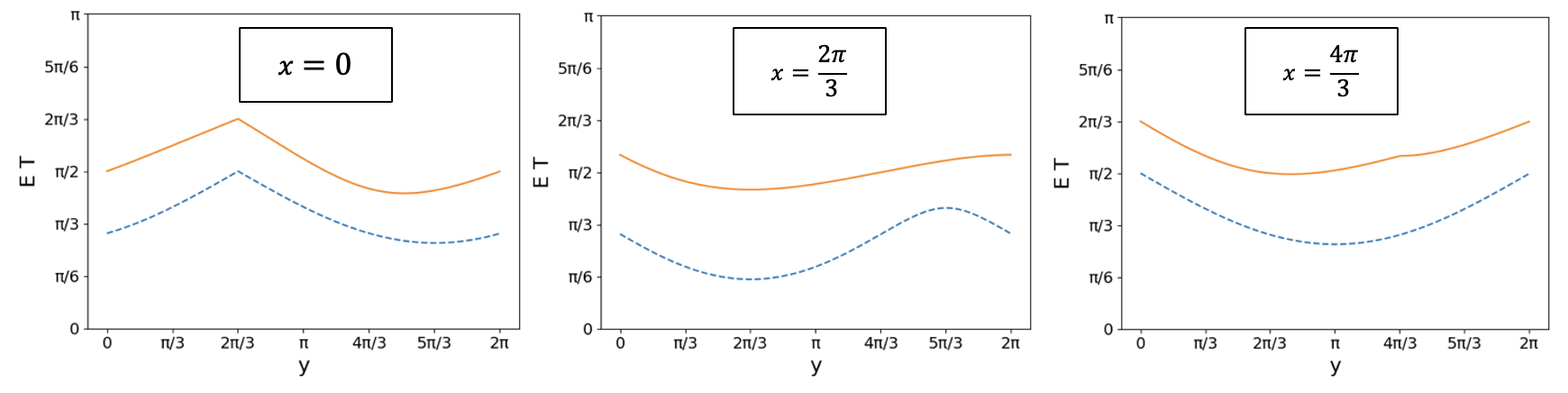}\vspace{0cm}
	\caption{The exact energy-time relation for the unitary $U_1$  as a function of $y$ for various values of $x$(continuous line) versus the time-energy bound (\ref{ineq1}) (dashed line). }
	\label{U1}
\end{figure}

\begin{figure}[H]
	\centering	\includegraphics[width=14cm]{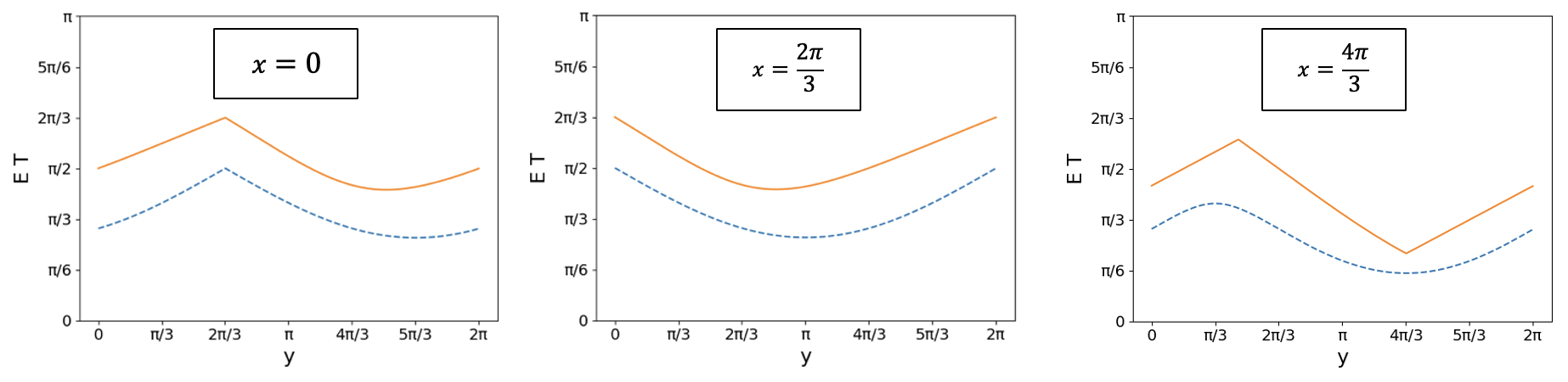}\vspace{0cm}
	\caption{The exact energy-time relation for the unitary $U_2$ as a function of $y$ for various values of $x$(continuous line) versus the time-energy bound (\ref{ineq1}) (dashed line).  }
	\label{U3}
\end{figure}

\ni In this way, one can restrict the transformations between MUB's which have only two free parameters, instead of five. Thus the two classes of unitaries are the following: 
\begin{equation}
	U_1(x,y)  =\frac{1}{\sqrt{3}} \begin{pmatrix}
		1& 	 e^{ix }& 	 e^{iy}\\
		1& 	 \overline{\omega} e^{ix} & 	 \omega   e^{iy} \\
		1 &	  \omega e^{ix }&	 \overline{\omega}  e^{iy}	\end{pmatrix}
\end{equation}
\ni and
\begin{equation}
	U_2 (x,y) =\frac{1}{\sqrt{3}} \begin{pmatrix}
		1& 	 e^{ix }& 	 e^{iy}\\
		1& 	  \omega e^{ix} & 	\overline{ \omega}  e^{iy } \\
		1 &	  \overline{\omega} e^{ix }&	 \omega  e^{iy }
	\end{pmatrix}
\end{equation}
where $x$ and $y$ are real parameters. These two manifolds are disjoint, figure (\ref{UUU}). Consider the class $U_1$.  We can determine the energy-time relation for this unitary by numerically evaluating its eigenvalues,  leads to energy times eigenvalues in the form
$E_0T, E_1T $ and $E_2T$. Note that, due to the ambiguity in taking the logarithm of $U$, even after choosing the energies to be as low as possible, several different time-energy relations follow. (see the remark in the introduction).  We have chosen the order of energies in such a way to lead to the minimum time-energy relation.  Figure (\ref{U1}) shows the exact energy-time relation for the unitary $U_1$ as a function of $y$ for various values of $x$. Figure (\ref{U3}) shows the same relation for the unitary $U_2$. On the same figures, the bound (\ref{ineq1}) is also plotted. This plot is obtained simply by evaluating $|tr(U_1)|=|tr(U_2)|=\frac{1}{\sqrt{3}}|1+{\omega}(e^{ix}+e^{iy})|$ and using (\ref{ineq1}). It is evident that the exact calculation (which is obtained after numerical work) validates the bound (\ref{ineq1}) which is obtained by a very simple calculation. 

\section{Conclusion and outlook}

We have established bounds on the speed limit of quantum evolution by unitary operators in arbitrary dimensions. These time-energy bounds do not depend on the initial and final state but depend only on the trace of the  unitary operator which is to be implemented and the gross characteristics of the  energy spectrum of the Hamiltonian which generates this unitary evolution. To this end, we have shown how one can circumvent the ambiguities encountered in taking the logarithm of a unitary operator $U=e^{-iHT}$. 
The bounds that we find can be thought of as the generalization of the Mandelstam-Tamm (MT) bound, the Margolus-Levitin (ML) bound and the dual Margolus-Levitin (dual ML) bounds to the realm of unitary operators. After proving these bounds, we have studied several classes of transformations, including change of basis between MUB's, permutations, Fourier transform, and the Grover operation. We have also explicitly studied the general transformation for qubits and MUBs transformation for qutrits. When applicable, we have compared our bounds with those obtained in previous works.\\

\noindent {\Large{\bf Acknowledgment:}} This research was supported in part by Iran National Science Foundation, under Grant No.4022322.  The authors wish to thank Vahid  Jannessary,  Shayan Roofeh and Amirhossein Tangestani for their valuable comments and suggestions. 

\newpage
\bibliography{citations.bib}
\end{document}